# uFLIP: Understanding Flash IO Patterns

**Luc Bouganim**
INRIA Rocquencourt and University of Versailles
FRANCE
Luc.Bouganim@inria.fr

**Björn Þór Jónsson**
Reykjavík University
ICELAND
bjorn@ru.is

**Philippe Bonnet**
University of Copenhagen
DENMARK
bonnet@diku.dk

## ABSTRACT
Does the advent of flash devices constitute a radical change for secondary storage? How should database systems adapt to this new form of secondary storage? Before we can answer these questions, we need to fully understand the performance characteristics of flash devices. More specifically, we want to establish what kind of IOs should be favored (or avoided) when designing algorithms and architectures for flash-based systems. In this paper, we focus on flash IO patterns, that capture relevant distribution of IOs in time and space, and our goal is to quantify their performance. We define uFLIP, a benchmark for measuring the response time of flash IO patterns. We also present a benchmarking methodology which takes into account the particular characteristics of flash devices. Finally, we present the results obtained by measuring eleven flash devices, and derive a set of design hints that should drive the development of flash-based systems on current devices.

## 1. INTRODUCTION
*Tape is dead, disk is tape, flash is disk* [5]. The flash devices that now emerge as a replacement for mechanical disks are complex devices composed of flash chip(s), controller hardware, and proprietary software that together provide a block device interface via a standard interconnect (e.g., USB, IDE, SATA). Does the advent of such flash devices constitute a radical departure from hard drives? Should the design of database systems be revisited to accommodate flash devices? Must new systems be designed differently to take full advantage of flash device characteristics?

In trying to answer these questions, a tempting short-cut is to assume that flash devices behave as the flash chips they contain. Flash chips are indeed very precisely specified, they have interesting properties (e.g., read/program/erase operations, no updates in-place, random reads equivalent to sequential reads), and many researchers have used their characteristics to design new algorithms [8][11][14]. The problem is, however, that commercially available flash devices *do not behave as flash chips*. They provide a block interface, where data is read and written in fixed sized blocks and integrate layers of software that manage block mapping, wear-leveling and error correction. As a consequence, flash devices do not provide explicit erase operations, and there is, a priori, no reason to avoid in-place updates. In terms of performance, flash devices are also much more complex than flash chips. For instance, block writes directed to the flash devices are mapped to program and erase operations at different granularities and as a result the performance of writes is not uniform in time. It would therefore be a mistake to model flash devices as flash chips.

So, how can we model flash devices? The answer is not straightforward because flash devices are both complex and undocumented. They are black boxes from a system's point of view.

### 1.1 Understanding Flash Devices
A first step towards the modeling of flash devices is to have a clear and comprehensive understanding of their performance. The key issue is to determine the kinds of IOs (or IO sequences) that should be favored (or avoided) when designing algorithms and architectures for flash-based systems.

In order to study this issue, we need a benchmark that quantifies the performance of flash devices. By applying such a benchmark to current and future devices, we can start making progress towards a comprehensive understanding. While individual devices are likely to differ to some extent, the benchmark should reveal common behaviors that will form a solid foundation for algorithm and system design. In this paper, we propose such a benchmark.

Defining a benchmark for flash devices is not a trivial task, however. Since the behavior of flash devices is determined by layers of undocumented software, we cannot make any safe assumptions. This has three main consequences. First, to capture the performance characteristics we must define a broad benchmark that casts light on all relevant usage patterns. Second, since the space of relevant usage patterns is large, we must focus on simple measurements that are easily analyzed. Last, the complex behavior calls for sound benchmarking methodology.

Recently, there has been increasing awareness of the importance of benchmarking activities in general, and benchmarking methodologies in particular (e.g., see [2][13]). For example, it has been demonstrated that incorrect benchmarking methodology may lead to distorted benchmark results [2][13]. In this paper, we therefore put significant emphasis on devising a sound benchmarking methodology for flash devices.

## 1.2 Related Work

So far, only a handful of papers have attempted to understand the overall performance of flash devices. Lee et al. focused on benchmarking SSD performance for typical database access patterns, but used only a single device for their measurements [9]. Myers measured the performance of database workloads over two flash devices [10]. In comparison, our benchmark is not specific to database systems. We study a variety of IO patterns, defined as the distribution of IOs in time and space.

Ajwani et al. analyzed the performance of a large number of flash devices, but using ad-hoc methodology [1]. By contrast, we identify benchmarking methodology as a major challenge. It is indeed very easy to get meaningless results when measuring a flash device because of the non-uniform nature of writes. Huang et al. attempted an analysis of flash device behavior, but neither proposed a complete methodology nor made measurements [7]. There are thus no relevant flash device benchmarks.

Many existing benchmarks aim to measure disk performance (see [13] for an excellent survey and critique). None of those benchmarks, however, accounts for the non-uniform performance of writes that characterizes flash devices.

## 1.3 Contributions of the Paper

This paper makes the following three major contributions:

1. We define the uFLIP benchmark (Section 3), a component benchmark for understanding flash device performance. uFLIP is a collection of nine micro-benchmarks defined over IO patterns.

2. We define a benchmarking methodology (Section 4) that accounts for the complex and non-uniform performance of flash devices.

3. We apply the uFLIP benchmark to a set of eleven flash devices (Section 5), ranging from low-end to high-end devices. Based on our results, we discuss a set of design hints that should drive the development of flash-based systems on current devices.

We believe that the investigation of flash device behavior deserves strong and continuous effort from the research community; an effort that we instigate in this paper. Therefore, the uFLIP software and the detailed results (tens of millions of data points) are available on a web site (www.uflip.org) that we expect to be used and completed by the community.

# 2. FLASH DEVICES

The uFLIP benchmark is focused on flash devices—such as solid state disks (SSDs), USB flash drives, or SD cards—which are packaged as block devices. As Figure 1 below illustrates, flash devices contain flash chips and controllers whose role is to provide the block abstraction at the flash device interface.

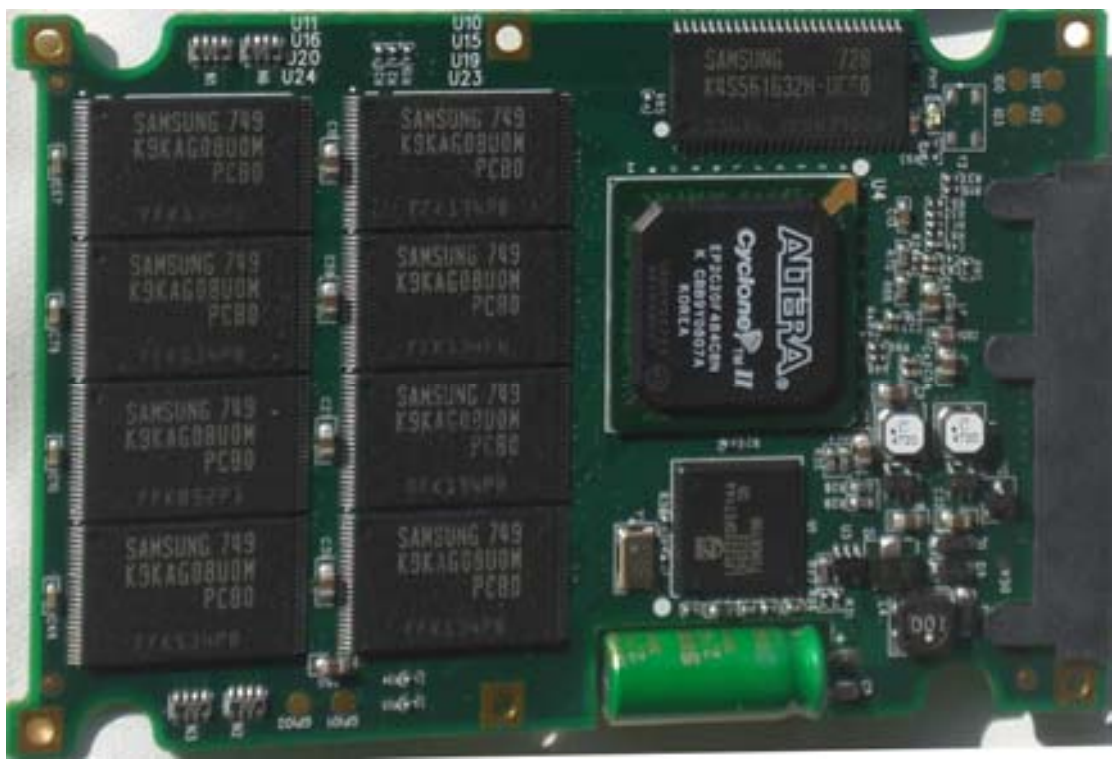

**Figure 1: Internal view of the Memoright 32 GB SSD**

While the details of flash devices vary significantly, there are certain common traits in the architecture of the flash chips and of the block manager that impact their performance and justify our focus on IO patterns [3]. In this section we review those common traits.

## 2.1 Flash Chips

A flash chip is a form of EEPROM (Electrically Erasable Programmable Read-Only Memory), where data is stored in independent arrays of memory cells. Each array is a *flash block*, and rows of memory cells are *flash pages*. Flash pages may furthermore be broken up into *flash sectors*.

Each memory cell stores 1 bit in single-level cell (SLC) chips, or 2 or more bits in multi-level cell (MLC) chips. MLC chips are both smaller and cheaper, but they are slower and have a shorter expected life span. By default each bit has the value 1. It must be *programmed* to take the value 0 and *erased* to go back to value 1. Thus, the basic operations on a flash chip are read, program and erase, rather than read and write.

Flash devices designed for secondary storage are all based on NAND flash, where the rows of cells are coupled serially, meaning that data can only be read and programmed at the granularity of flash pages (or flash sectors).Writes are performed one page (or sector) at a time, and sequentially within a flash block in order to minimize write errors resulting from the electrical side effects of writing a series of cells.

Erase operations are only performed at the granularity of a flash block (typically 64 flash pages). This is a major constraint that the block manager must take into account when mapping writes onto program and erase commands. Most flash chips can only support up to $10^5$ erase operations per flash block for MLC chips, and up to $10^6$ in the case of SLC chips. As a result, the block manager must implement some form of wear-leveling to distribute the erase operations across blocks and increase the life span of the device. To maintain data integrity, bad cells and worn-out cells are tracked and accounted for. Typically, flash pages contain 2KB of data and a 64 byte area for error correcting code and other bookkeeping information.

Modern flash chips can be composed of two planes, one for even blocks, the other for odd blocks. Each flash chip may contain a page cache. The block manager should leverage these forms of parallelism to improve performance.

## 2.2 Block Manager

In all flash devices, the core data structures of the block manager are two maps between blocks, represented by their *logical block addresses* (LBAs), and flash pages. A *direct* map from LBAs to flash pages is stored on flash and in RAM to speed up reads, and an *inverse* map is stored on flash, to re-build the direct map during recovery. There is a trade-off between the improved read performance due to the direct map and degraded write performance due to the update of the inverse map (updates of bookkeeping information for a page may cause an erase of an entire block).

The software layer responsible for managing these maps both in RAM (inside the micro-controller that runs the block manager) and on flash is called *flash translation layer* (FTL). Using the direct map, the FTL introduces a level of indirection that allows trading expensive writes-in-place (with the erase they incur) for cheaper writes onto free flash pages.

Each update on a free flash page, however, leaves an obsolete flash page (that contains the before image). Over time such obsolete flash pages accumulate, and must subsequently be reclaimed synchronously or asynchronously. As a result, we must assume that the cost of writes is not homogeneous in time (regardless of the actual reclamation policy). Some block writes will result in flash page writes with a minimum bookkeeping overhead, while other block writes will trigger some form of page reclamation and the associated erase. Assuming a flash device contains enough RAM and autonomous power, the flash translation layer might be able to cache and destage both data and bookkeeping information.

The density of NAND flash chips is doubling every year [12]. As a result, the capacity of flash devices increases exponentially. This has a direct impact on the size of the direct map from LBAs to flash pages. If the direct map does not fit in RAM, then the cost of reads will also become non-uniform in time as portions of the map will need to be swapped in to complete a look-up. Conversely, the increase in storage capacity has no direct impact on the distribution of expensive writes as obsolete pages still must be reclaimed to guarantee constant flash capacity. Flash device constructors might be able to use a larger fraction of the storage space for bookkeeping, which can mitigate the need to reclaim pages.

## 2.3 Device State and IO Patterns

While the principles of the flash translation layer described above are well known, the details of the design decisions made for a given flash device, and the associated performance trade-offs, are typically not documented: Flash devices are black-boxes.

Since the physical layout of data on flash devices is stored and manipulated via the direct map data structure, which manages the devices at some fixed granularity, there is no direct correspondence between an arbitrary IO request to the flash device and its translation to a physical request to a flash chip. Instead, the physical request is based in a complex manner on the current state of the direct map, which in turn is based on the entire history of previous IO requests.

This circular dependency makes benchmarking flash devices particularly hard. The main factors that impact performance are thus the device *state* and the distribution of incoming IO requests in space (i.e., their LBA) and time, or *IO patterns*. The goal of the uFLIP benchmark is to characterize the relevant IO patterns and, given a well defined device state, how they impact performance.

# 3. THE uFLIP BENCHMARK

In this section we propose uFLIP, a new benchmark for observing and understanding the performance of flash devices.

The uFLIP benchmark is a set of micro-benchmarks based on IO patterns. In theory, IO patterns can be arbitrarily complex. In uFLIP, we focus on a set of potentially relevant IO patterns defined through a small set of baseline patterns, functions and parameters. These IO patterns are described in Section 3.1.

The IO patterns thus defined still cover a very large design space, some of which is not useful for analysis. We have therefore used three design principles to further narrow the uFLIP benchmark to a set of nine micro-benchmarks. The details of these nine micro-benchmarks are given in Section 3.2.

We believe that the nine uFLIP micro-benchmarks together capture the characteristics of flash devices quite well. In Section 3.3 we further argue that they form a benchmark that fulfils key quality criteria defined in the literature.

## 3.1 Defining IO Patterns

The basic construct of uFLIP is an IO pattern, which is simply a sequence of IOs. In each pattern, we refer to the $i^{th}$ submitted IO as $IO_i$. Each IO is (as usual) defined by four attributes:

1. $t(IO_i)$: the time at which the IO is submitted;
2. $IOSize(IO_i)$: the IO size;
3. $LBA(IO_i)$: the IO location or logical block address;
4. $Mode(IO_i)$: the IO mode, which is either read or write.

In theory, IO patterns can be arbitrarily complex, as arbitrary functions can be used to generate the four attributes. We define four *baseline patterns* as sequential reads, sequential writes, random reads, and random writes for consecutive IOs of a given size; these are the patterns typically used in practice. In order to increase the range of relevant patterns for our benchmark, however, we introduce a few simple parameterized functions:

- The time, $t(IO_i)$, is defined through one of three different functions: a) *consecutive*, where $IO_{i+1}$ starts as soon as $IO_i$ finishes, as in the baseline patterns; b) *pause*(*Pause*), where a pause of length *Pause* is introduced in between all IOs; or c) *burst*(*Pause*, *Burst*), where a pause of length *Pause* is introduced between groups of *Burst* IOs.

  Note that both the *pause* and *consecutive* functions can be defined using the *burst* function, as *pause*(*p*) = *burst*(1, *p*) and *consecutive* = *burst*(0, –). We feel, however, that the *pause* and *consecutive* functions are important enough to be considered separately.

- $IOSize(IO_i)$ is simply defined as the identity function over the parameter *IOSize*.
- The location of the IO, $LBA(IO_i)$ is defined through one of four different functions: a) *sequential*; b) *random*; c) *ordered(Incr)*, where a linear increment (or decrement) is applied to each LBA in the pattern, determined by the linear coefficient *Incr*; or d) *partitioned*(*Partitions*), where we divide the target space into *Partitions* partitions which are considered in a round robin fashion; within each partition IOs are sequential.

For each of these functions, the address is first computed assuming an alignment to *IOSize* boundaries, and relative to a specified target location on the flash device (*TargetOffset*). Additionally, we must specify the size of the target space of the IO pattern (*TargetSize*) and whether the IO is indeed aligned or not (*IOShift*).

- Lastly, *Mode*($IO_i$) is a constant function yielding two values, *read* or *write*.

Figure 2 illustrates the use of many of these functions and parameters. Figure 2.a illustrates the alignment of the *LBA* function, the impact of the *IOSize* and *IOShift* parameters, and the response time of the IO, *rt*($IO_i$). Figure 2.b illustrates the impact of the *TargetOffset* and *TargetSize* parameters, as well as the *Partitions* parameter. Figure 2.c, on the other hand, illustrates the impact of the *Incr*, *Pause*, and *Burst* parameters for a sequential pattern. Note that for each pattern, we must also specify its length (*IOCount*) and warm-up period (*IOIgnore*). Setting the *TargetOffset*, *IOIgnore*, and *IOCount* parameters is part of the benchmarking methodology described in Section 4.

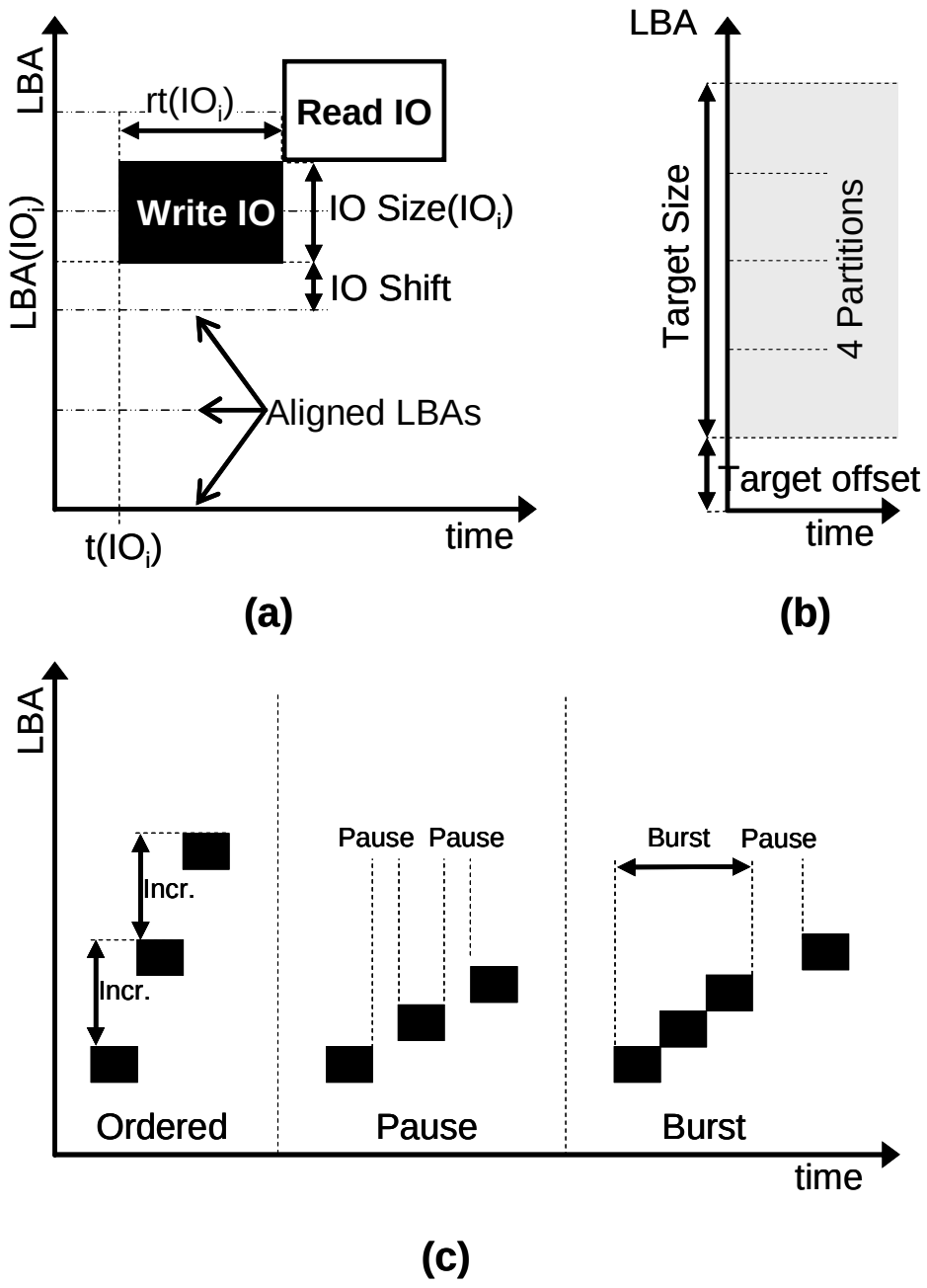

**Figure 2: Parameters and functions**

Based on these functions, we consider three different types of potentially relevant IO patterns:

- **Basic patterns** obtained by selecting one function for each of the four dimensions, assigning values to all the parameters. To give an idea of the size of the pattern space covered, there are 3×1×4×2 =24 basic IO patterns.
- **Mixed patterns** obtained by combining basic patterns. Additional parameters are then required to control the mix of patterns. Even considering only combinations of two basic patterns results in a very large space of IO patterns, with one additional parameter required to control the mix.
- **Parallel patterns** obtained by replicating a single basic pattern with a given degree of parallelism (*ParallelDegree*) or by mixing, in parallel, different basic patterns.

## 3.2 The uFLIP Micro-Benchmarks

Even with the restrictions that we have introduced, the space of potentially relevant IO patterns is far too large to be explored exhaustively. To further reduce the complexity of the benchmark, we adopt the following design principles:

1. An execution of a reference pattern against a device is called a run. For each run, we measure and record the response time for individual IOs[1] and compute statistics (min, max, mean, standard deviation) to summarize it.

   A collection of runs of the same reference pattern is called an experiment. To enable sound analysis of an experiment's results (set of computed statistics), we design each experiment around a single varying parameter.[2]

2. We then define a *micro-benchmark* as a collection of related experiments over the baseline patterns. The experiments of the same micro-benchmark thus have the same varying parameter, and the same value for all other functions and parameters. This gives a total of nine micro-benchmarks, as follows:

   Considering the basic pattern parameters (one of *IOSize*, *IOShift*, *TargetSize*, *Partitions*, *Incr*, *Pause*, or *Burst*) yields seven micro-benchmarks.

   Since mixed patterns need parameters to control the mix, restricting to varying a single parameter allows us to mix only the baseline patterns. This leads to one more micro-benchmark with six combinations of baseline patterns, controlled by a *Ratio* parameter.

   Similarly, in order to focus only on one parameter, *ParallelDegree*, for parallel patterns, we choose to replicate only the four baseline patterns to implement parallelism.

3. Each micro-benchmark is based on the four baseline patterns, departing from the baseline patterns only to accommodate the particular parameter being varied.

We now define the nine uFLIP micro-benchmarks by describing informally the sets of reference patterns and the parameter that is varied. Table 1 fully specifies every micro-benchmark with a formula for the four IO attributes, as well as the name and range of the relevant parameters. In the table, we first give the formulas for the baseline patterns and then focus on the changes required to accommodate the varying parameter in each case. We present the micro-benchmarks by first considering location parameters, then parallel and mixed patterns, and end with the timing parameters.

1. **Granularity (*IOSize*):** The flash translation layer manages a direct map between blocks and flash pages, but the granularity at which this mapping takes place is not documented. The *IOSize* parameter allows determining whether a flash device favors a given granularity of IOs.
2. **Alignment (*IOShift*):** Using a fixed *IOSize* (e.g., chosen based on the first micro-benchmark), we study the impact of alignment on the baseline patterns by introducing the *IOShift* parameter and varying it from 0 to *IOSize*.

---

[1] One could consider other metrics such as space occupation or aging. Given the block abstraction the only way to measure space occupation is indirectly through write performance measurements. Measuring aging is difficult since reaching the erase limit (with wear leveling) may take years. Measuring power consumption, however, should be considered in future work.

[2] Considering more than one varying parameter would lead to multi-dimensional graphs, which are too complex to analyze.

**Table 1: Micro-Benchmark definitions**

| Micro-benchmark | Attribute | Pattern Definitions | Examples |
|---|---|---|---|
| **Baseline patterns** | $t(IO_i)$<br>$IOSize(IO_i)$<br>$LBA(IO_i)$<br><br>$Mode(IO_i)$ | *consecutive:* $t(IO_{i-1}) + rt(IO_{i-1})$<br>constant (32 KB in our experiments)<br>Rnd: $TargetOffset + random(TargetSize/IOSize) \times IOSize$<br>Seq: $TargetOffset + i \times IOSize$<br>read \| write | 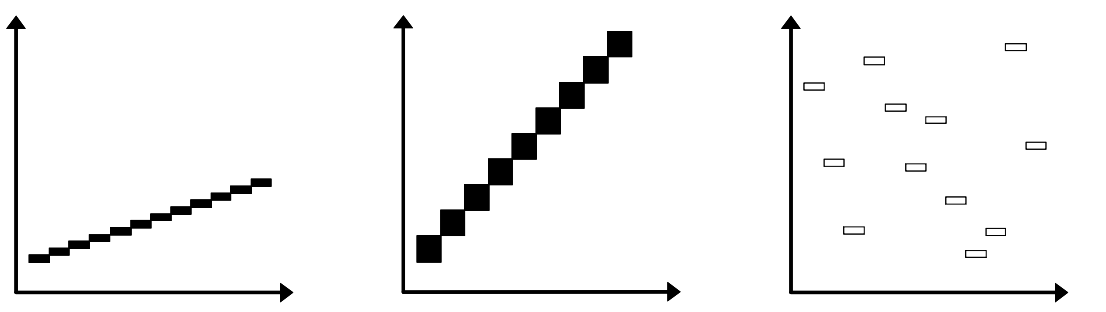<br>SW (*IOSize* = 8 KB) SW (*IOSize* = 32 KB) RR (*IOSize* = 8 KB) |
| **Granularity** | $IOSize(IO_i)$<br>$IOSize$ | $IOSize$<br>$[2^0, \ldots, 2^9] \times 512B$ (plus some non-powers of 2) | |
| **Alignment** | $LBA(IO_i)$<br><br><br>$IOShift$ | Rnd: $TargetOffset + IOShift +$<br>$random(TargetSize/IOSize) \times IOSize$<br>Seq: $TargetOffset + IOShift + i \times IOSize$<br>$[2^0, \ldots, IOSize/512] \times 512B$ | 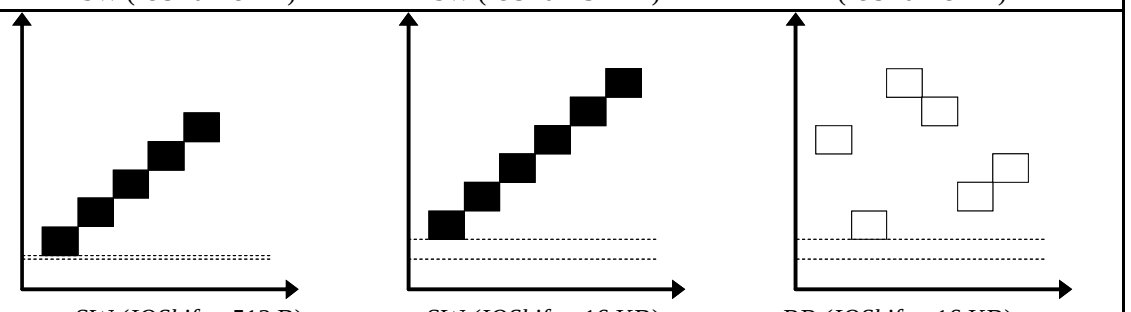<br>SW (*IOShift* = 512 B) SW (*IOShift* = 16 KB) RR (*IOShift* = 16 KB) |
| **Locality** | $LBA(IO_i)$<br><br>$TargetSize$ | Rnd: Unchanged<br>Seq: $TargetOffset + (i \times IOSize) \bmod TargetSize$<br>Rnd: $[2^0, \ldots, 2^{16}] \times IOSize$<br>Seq: $[2^0, \ldots, 2^8] \times IOSize$ | 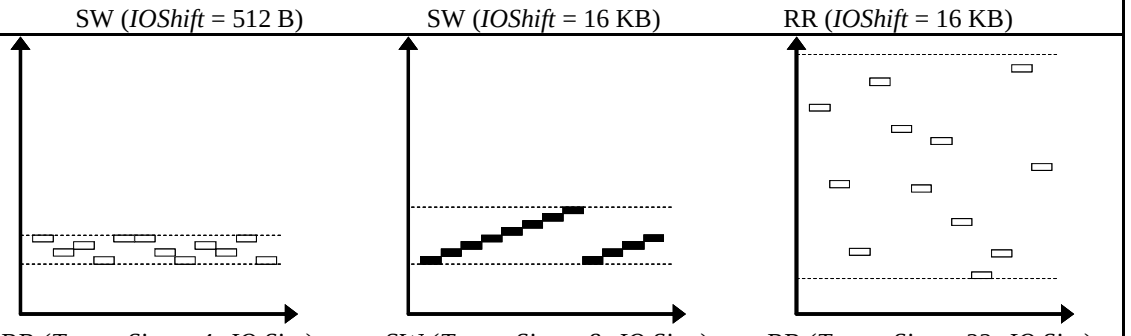<br>RR (*Target Size = 4×IO Size*) SW (*Target Size = 8×IO Size*) RR (*Target Size = 32×IO Size*) |
| **Partitioning** | $LBA(IO_i)$<br><br><br><br><br>$Partitions$ | Sequential patterns only<br>Seq: $P_i \times PS + O_i$ where<br>$PS = TargetSize/Partitions$<br>$P_i = i \bmod Partitions$<br>$O_i = \lfloor (i/Partitions) \times IOSize \rfloor \bmod PS$<br>$[2^0, \ldots, 2^8]$ | 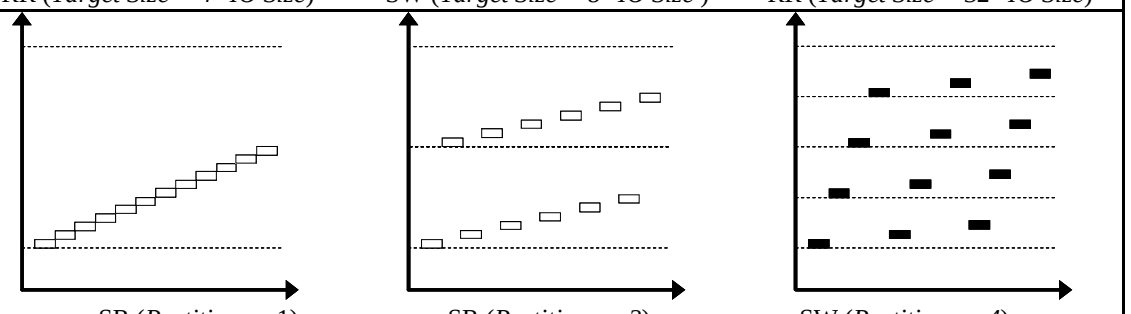<br>SR (*Partitions* = 1) SR (*Partitions = 2*) SW (*Partitions* = 4) |
| **Order** | $LBA(IO_i)$<br><br>$Incr$ | Sequential patterns only<br>Seq: $TargetOffset + (Incr \times i \times IOSize)$<br>$[-1, 0, 2^0, \ldots, 2^8]$ | 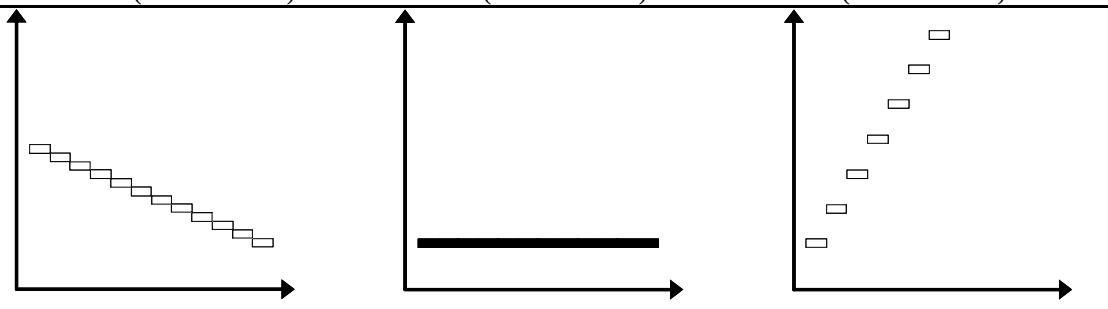<br>SR (*Incr* = –1 ) SW (*Incr* = 0) SR (*Incr* = 4) |
| **Parallelism** | $LBA(IO_i)$<br><br><br>$ParallelDegree$ | *ParallelDegree* concurrent processes. For process *p*,<br>$TargetOffset_p = p \times TargetSize / ParallelDegree$<br>$TargetSize_p = TargetSize / ParallelDegree$<br>$[2^0, \ldots, 2^4]$ | 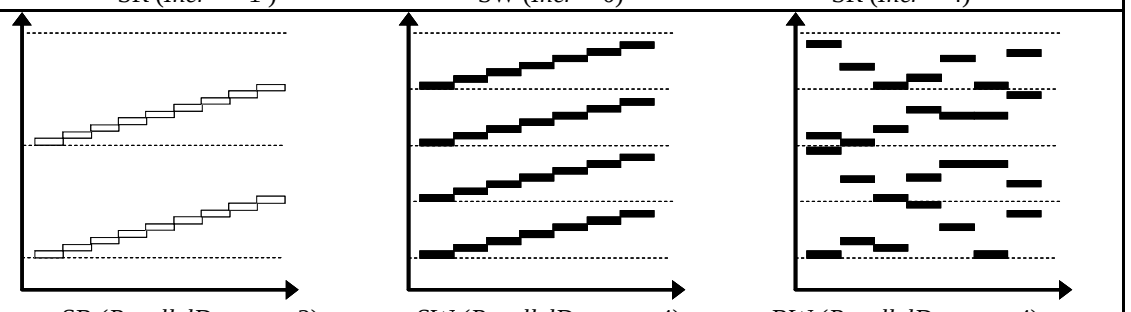<br>SR (*ParallelDegree* = 2) SW (*ParallelDegree* = 4) RW (*ParallelDegree* = 4) |
| **Mix** | *Pattern #1*<br>*Pattern #2*<br>*Ratio (#1/#2)* | SR SR SR RR RR SW<br>RR RW SW SW RW RW<br>$[2^0, \ldots, 2^6]$ | 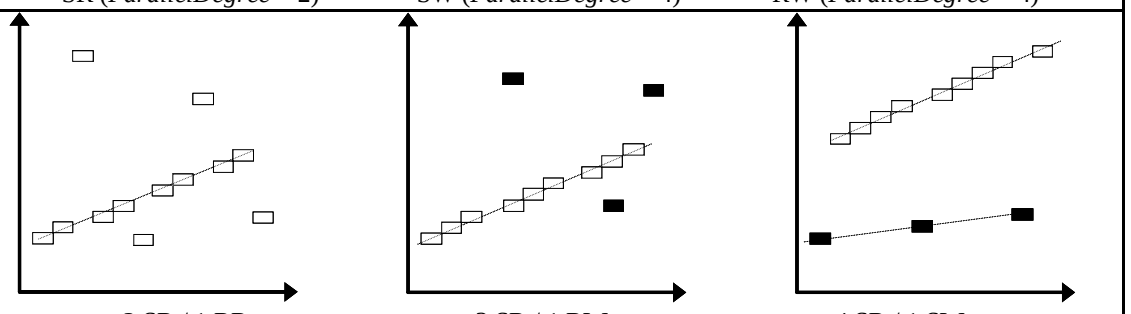<br>2 SR / 1 RR 3 SR / 1 RW 4 SR / 1 SW<br>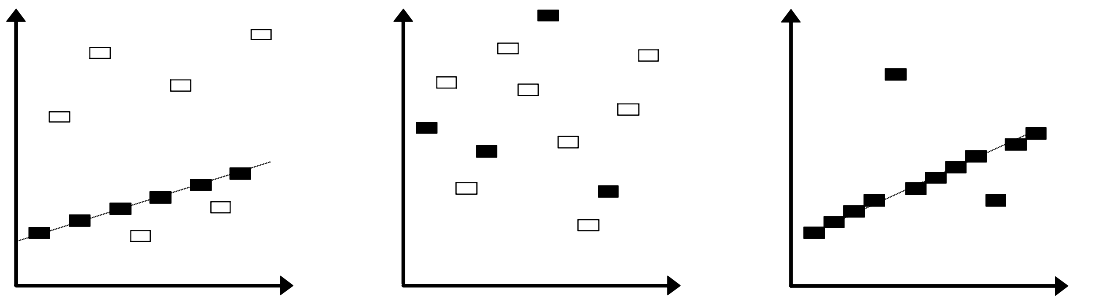<br>1 RR / 1 SW 2 RR / 1 RW 4 SW / 1 RW |
| **Pause** | $t(IO_i)$<br>$Pause$ | $t(IO_{i-1}) + rt(IO_{i-1}) + Pause$<br>$[2^0, \ldots, 2^8] \times 0.1$ msec | 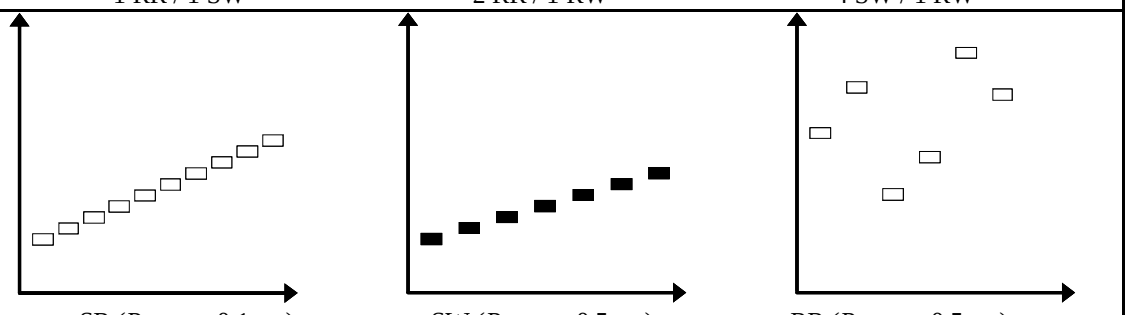<br>SR (*Pause* = 0.1 ms) SW (*Pause* = 0.5 ms) RR (*Pause* = 0.5 ms) |
| **Bursts** | $t(IO_i)$<br>$Pause$<br>$Burst$ | $t(IO_{i-1}) + rt(IO_{i-1}) + (i \bmod Burst) \times Pause$<br>e.g., 100 ms<br>$[2^0, \ldots, 2^6] \times 10$ | 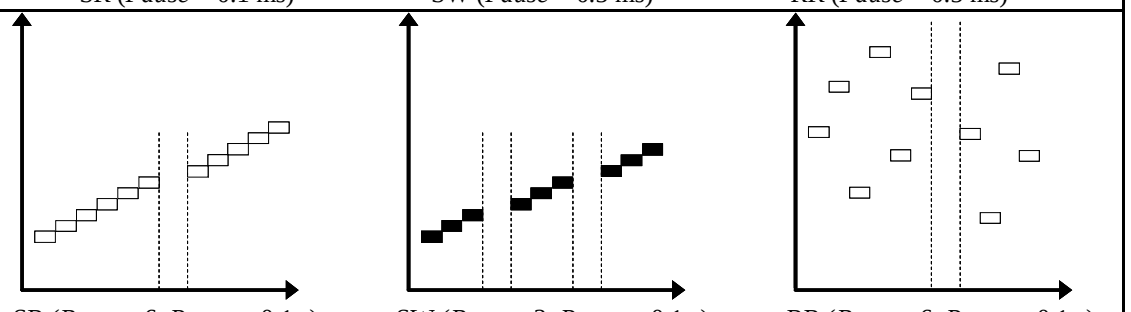<br>SR (*Burst* = 6, *Pause* = 0.1 s) SW (*Burst* = 3, *Pause* = 0.1 s) RR (*Burst* = 6, *Pause* = 0.1 s) |

3. **Locality (*TargetSize*):** We study the impact of locality of the baseline patterns, by varying *TargetSize* down to *IOSize*.

4. **Partitioning (*Partitions*):** The partitioned patterns are a variation of the sequential baseline patterns. We divide the target space into *Partitions* partitions which are considered in a round robin fashion; within each partition IOs are performed sequentially. This pattern represents, for instance, a merge operation of several buckets during external sort.

5. **Order (*Incr*):** The order patterns are another variation on the sequential patterns, where logical blocks are addressed in a given order. For the sake of simplicity, we consider a linear increase (or decrease) in the LBAs addressed in the pattern, determined by a linear coefficient *Incr*. We can thus define a) patterns with increasing LBAs (*Incr* > 1) or decreasing LBAs (*Incr* < 0), or b) in-place patterns (*Incr* = 0) where the LBA remains the same throughout the pattern.

   These mapping are simple, yet important and representative of different algorithmic choices: for example, a reverse pattern (*Incr* = –1) represents a data structure accessed in reverse order when reading or writing, the in-place pattern is a pathological pattern for flash chips, while an increasing LBA pattern represents the manipulation of a pre-allocated array, filled by columns or lines.

6. **Parallelism (*ParallelDegree*):** Since flash devices include many flash chips (even USB flash drives typically contain two flash chips), we want to study how they support overlapping IOs. We divide the target space into *ParallelDegree* subsets, each one accessed by a process executing the same baseline pattern. We vary the parameter *ParallelDegree* to study how well the flash device supports parallelism, and thus how asynchronous IO should be scheduled and how parallelism should be managed.

7. **Mix (*Ratio*):** We compose any two baseline patterns, for a total of six combinations. We vary the ratio to study how such mixes differ from the baselines.

8. **Pause (*Pause*):** This is a variation of the baseline patterns, where IOs are not contiguous in time. We use the *pause* function and vary the *Pause* parameter to observe whether potential asynchronous operations from the flash device block manager impact performance.

9. **Bursts (*Burst*):** This is a variation of the previous micro-benchmark, where the *Pause* parameter is set to a fixed length (e.g. 100 msec). The *Burst* parameter is then varied to study how potential asynchronous overhead accumulates in time.

## 3.3 Discussion

Even though uFLIP is not a domain-specific benchmark, it should still fulfill the four key criteria defined in the Benchmarking Handbook: portability, scalability, relevance and simplicity [6].

Because uFLIP defines how IOs should be submitted, uFLIP has no adherence to any machine architecture, operating system or programming language: uFLIP is portable.

Also, uFLIP does not depend on the form factor of flash device being studied, and we have indeed run uFLIP on USB flash drives, SD cards, IDE flashes and SSD drives: uFLIP is scalable.

We believe uFLIP is relevant for algorithm, system and flash designers because the nine micro-benchmarks reflect flash device characteristics as well as the characteristics of the software that generates IOs. It is neither designed to support decision making nor to reverse engineer flash devices.

Whether uFLIP satisfies the criteria of simplicity is debatable. The benchmark definition itself is quite simple, and indeed we have reduced an infinite space of IO patterns down to nine micro-benchmarks that define how IOs are submitted using very simple formulas and parameters.

On the other hand, our decision to measure response time for each submitted IO means that the benchmark results are very large and analyzing those results is not straightforward. We felt, however, that requiring such analysis is fundamental to achieving our goal of understanding the flash device performance. We have therefore designed a visualization tool that facilitates interactive result analysis.

Furthermore, benchmarking flash devices is inherently far from simple, e.g., due to the impact of the device state on the performance of individual operations. The methodology we present in the next section addresses this issue.

In summary, we believe that uFLIP is as simple as possible, given the complexities of flash device benchmarking, and that further simplifications would lead to loss of understanding. We note that we have achieved this simplicity by following strictly the three design principles outlined in Section 3.2.

# 4. BENCHMARKING METHODOLOGY

Measuring flash device performance is very challenging. First, as discussed in Section 2, the state of the device impacts its performance. Second, because response time is not uniform, each experiment must be long enough to capture the performance variations of the device under study. Third, consecutive micro-benchmark runs should not interfere with each other. In the remainder of this section, we discuss these challenges in detail.

## 4.1 Device State

In our experiments, we have observed that ignoring the state of a flash device can lead to meaningless performance measurements; we now describe the most striking example. Out-of-the-box, the Samsung SSD (see Section 5 for details) had excellent random write performance (around 1 msec for a 16KB random write, compared to around 8 msec for other SSDs). After randomly writing the entire 32GB of flash, however, the write performance decreased by almost an order of magnitude and became comparable to the other SSDs.

In order to obtain repeatable results we should run the micro-benchmarks from a well-defined initial state, which is independent of the complete IO history. Since flash devices only expose a block device API, however, we cannot erase all blocks and get back to factory settings. And, because flash devices are black boxes, we cannot know their exact state. We therefore make the following assumption for the uFLIP benchmark: *Writing the whole flash device completely yields a well-defined state*. The rationale is that following a complete write of the whole flash device, both the direct and indirect maps managed by the FTL are filled and well-defined.

We therefore propose to enforce an initial state for the benchmark by performing random IOs of random size (ranging from 0.5KB to the flash block size, 128 KB) on the whole device. The advantage of this method is that it is quite stable, as only sequential writes disturb the state significantly. In order to limit the impact of sequential writes, we direct them to distinct target spaces (specified by *TargetOffset*) when running the micro-benchmarks. The main disadvantage of our method is that it is slow, but since it is typically only done once this is acceptable.

The alternative, performing a complete rewrite of the device using sequential IOs of a given size, is faster but may be less stable for many devices. The reason for the increased instability is that random writes, badly aligned IOs, or IOs of different sizes, impact a sequential state much more significantly than a random state. Studying in detail the impact of the initial state on performance is, however, a topic for future work.

## 4.2 Start-up and Running Phases

Consider a device where the first 128 random writes are very cheap (400 µsec), and where the subsequent random writes oscillate between very cheap (400 µsec) and very expensive (27 msec). Now, say that we run the random write baseline pattern with *IOCount* = 512, which would seem to be long enough. In this case, the measured time will be about 25% lower than it should be; with shorter experiments the difference is even more pronounced. As another example, running the Mix micro-benchmark on Random Read and Write patterns with an *IOCount* of 512 will lead to entirely meaningless results when *Ratio* is higher than 4 (more than 4 reads for every write), as then our measurements only capture the initial, very cheap random writes. If we are not careful, in fact, we might even conclude that a read-mostly workload can absorb the cost of the random writes.

We propose a two-phase model to capture response time variations within the course of a micro-benchmark run. In the first phase, which we call *start-up phase*, response time is cheap. Such a start-up phase can occur when expensive operations are delayed, e.g., due to buffering or lazy garbage collection. In the second phase, which we call *running phase*, response time is typically oscillating between two or more values.

We thus characterize each device by two parameters: *start-up*, which defines the number of IOs for the start-up phase, and *period*, which defines the number of IOs in one oscillation in the running phase. In order to measure *start-up* and *period*, we run all four baseline patterns (SR, RR, SW and RW) with a very large *IOCount*. By plotting the IO costs, we can then identify the two phases for each pattern and derive upper bounds across the patterns for *start-up* and *period* (note that the start-up phase may not be present, in which case *start-up* = 0). Furthermore, we can determine the variability in the IO times.

The impact of this two-phase model on the benchmarking methodology is twofold. First, for each experiment we must adjust *IOCount* to capture both the start-up phase and the running phase (a sufficiently large number of periods). Second, we must ignore the start-up phase when summarizing the results of each run, so that we can use a statistical representation (min, max, mean, standard variation) to represent the response times obtained during the running phase. We therefore select *IOIgnore* as long enough to cover the start-up phase and *IOCount* as long enough to cover a sufficient number of periods to allow for convergence to the correct average response times.

Note that the value of *IOCount* has a direct impact on the time it takes to run an experiment and on the flash size involved in this experiment. As we saw in the previous sub-section, we must limit the portion of flash impacted by each experiment in order to avoid resetting the initial state too often during the course of benchmarking. Overestimating *IOCount* thus leads to a waste of time. Underestimating *IOCount*, on the other hand, leads to decreased precision and possibly incorrect results. Defining a method for automatically finding an appropriate *IOCount* value is a topic for future work.

Once *IOCount* is set to an appropriate value for each experiment, we define a benchmark plan that defines a sequence of state resets and micro-benchmarks, where those experiments involving sequential writes are delayed and grouped together in such a way that their allocated target space does not overlap, meaning that state resets are inserted only when the size of the accumulated target space involved in sequential write patterns is larger than the size of the flash device. Note that for the large flash devices (32 GB) the state is in fact never reset.

## 4.3 No Interference

Consecutive benchmark runs should not interfere with each other. Consider a device that implements an asynchronous page reclamation policy. Its effects should be captured in the running phase defined above. We must make sure, however, that the effect of the page reclamation triggered by a given run has no impact on subsequent, unrelated runs.

To evaluate the length of the pause between runs, we rely on the following experiment. We submit sequential reads, followed by a batch of random writes, and sequential reads again. We count the number of sequential reads in the second batch which are affected by the random writes. We then use this value as a lower bound on the pause between consecutive runs. Note that when benchmarking a device with unknown properties, this is only an educated guess, and therefore we propose to significantly overestimate the length of the pause.

Another type of potential interference is due to the file system, the operating system and the device drivers on the server hosting the flash device under study. Those layers of software introduce complexity and thus tend to complicate the analysis of the benchmark results. We thus use direct IO in order to bypass the host file system and synchronous IO to avoid the parallelism features of the operating system and device drivers.[3]

# 5. FLASH DEVICE EVALUATION

In this section, we report on our experimental evaluation of a range of flash devices, using the uFLIP benchmark. In Section 5.1 we describe our benchmark preparation following our methodology laid out in Section 4. In Section 5.2 we present and analyze the benchmark results. Finally, in Section 5.3, we discuss a set of design hints that can be drawn from our results.

[3] The lowest layer of the file system is the disk scheduler, which actually submits IO operations to the device driver. The disk scheduler is, as its name indicates, designed to optimize submission of IOs to disk. Whether disk schedulers should be redesigned for flash devices is an open question; the uFLIP benchmark should help in determining the answer.

## 5.1 Benchmark Preparation

We ran the uFLIP benchmark on an Intel Celeron 2.5GHz processor with 2GB of RAM running Windows XP. We ran each micro-benchmark using our own FlashIO software package (available at http://www.uflip.org/flashio.html). Each experiment was run three times. As the differences in performance were typically within 5%, we report the average of the three runs.

It was quite difficult to select a representative and diverse set of flash devices, as a) the flash device market is very active, b) products are not well documented (typically, random write performance is not provided!), and c) in fact, several products differ only by their packaging. We eventually selected the eleven different devices listed in Table 2, ranging from low-end USB flash drives or SD cards to high-end SSDs[4]. While we ran the entire uFLIP benchmark for all the devices, we only present results for seven representative devices indicated with an arrow in Table 1. Detailed information and measurements for all eleven flash devices can be found at http://www.uflip.org/results.html.

**Table 2: Selected flash devices**

| | Brand | Model | Type | Size | Price |
|---|---|---|---|---|---|
| → | Memoright | MR25.2-032S | SSD | 32 GB | $943 |
| | GSKILL | FS-25S2-32GB | SSD | 32 GB | $694 |
| → | Samsung | MCBQE32G5MPP | SSD | 32 GB | $517 |
| → | Mtron | SATA7035-016 | SSD | 16 GB | $407 |
| | Transcend | TS16GSSD25S-S | SSD | 16 GB | $250 |
| → | Transcend | TS32GSSD25S-M | SSD | 32 GB | $199 |
| → | Kingston | DT hyper X | USB drive | 8 GB | $153 |
| | Corsair | Flash Voyager GT | USB drive | 16 GB | $110 |
| → | Transcend | TS4GDOM40V-S | IDE module | 4 GB | $62 |
| → | Kingston | DTI 4GB | USB drive | 4 GB | $17 |
| | Kingston | SD 4GB | SD card | 2 GB | $12 |

**Random State Enforcement:** As prescribed in Section 4, we first filled each device with random writes of random size to enforce a random state. The time required for this varied significantly, ranging from 5 hours for the Memoright SSD to 35 days for the Corsair USB flash drive!

Although this is a significant time, it is still more efficient than enforcing a sequential state. Indeed, state enforcement is much faster with sequential state but the state also deteriorates faster as more workloads impact the state. Thus, the overall running time is longer with sequential state enforcement. In fact, sequential state enforcement on the Memoright SSD required a total formatting time of 17 hours, while a single format of 5 hours was sufficient for random state enforcement.

**Start-up and Running Phases:** As also prescribed in Section 4, we then ran the baseline patterns with large IOCount to measure *start-up* and *period* for each device. Figures 3 and 4 show two very representative traces from these measurements. In both figures, the *x*-axis shows the time in units of IO operations, while the *y*-axis shows the cost of each operation in msec (in logarithmic scale).

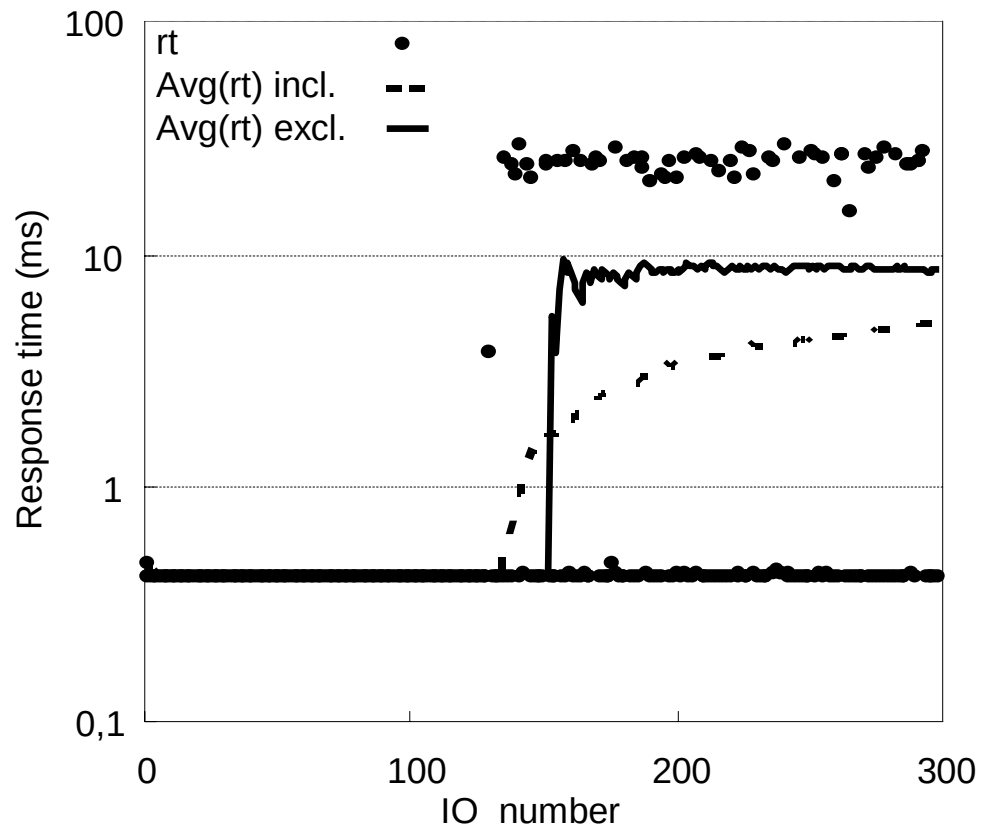

**Figure 3: Starting and running phase for Mtron SSD(RW)**

In Figure 3, which presents the RW baseline pattern for the Mtron SSD, we can easily distinguish between the start-up phase and the running phase. The start-up phase is about 125 IOs, while the period is quite short (tens of IOs). The dashed line represents the running average of response time, including the startup phase measurements, while the solid line represents the running average of response time, excluding the start-up phase measurements. As expected, excluding the start-up phase measurements resulted in a faster and more accurate representation of response time. In Figure 4, on the other hand, which presents the SW baseline pattern for the Kingston DTI USB flash drive, there is no startup phase and the period is about 128 operations.

With respect to start-up and running phases, we can basically divide the set of tested devices into two classes. The Memoright and Mtron SSDs both have a startup phase for random writes followed by oscillations with a very small period. They do not show startup for SR, RR and SW. For these devices, care should be taken when running experiments that involve a small number of RW, especially Mix patterns, since the startup phase should be scaled-up according to the number of RW IOs.

The other nine devices have no startup phase but show small oscillations for RR, larger ones for SW and sometimes large oscillations for RW (with some impressive variations between 0.25 and 300 msec).

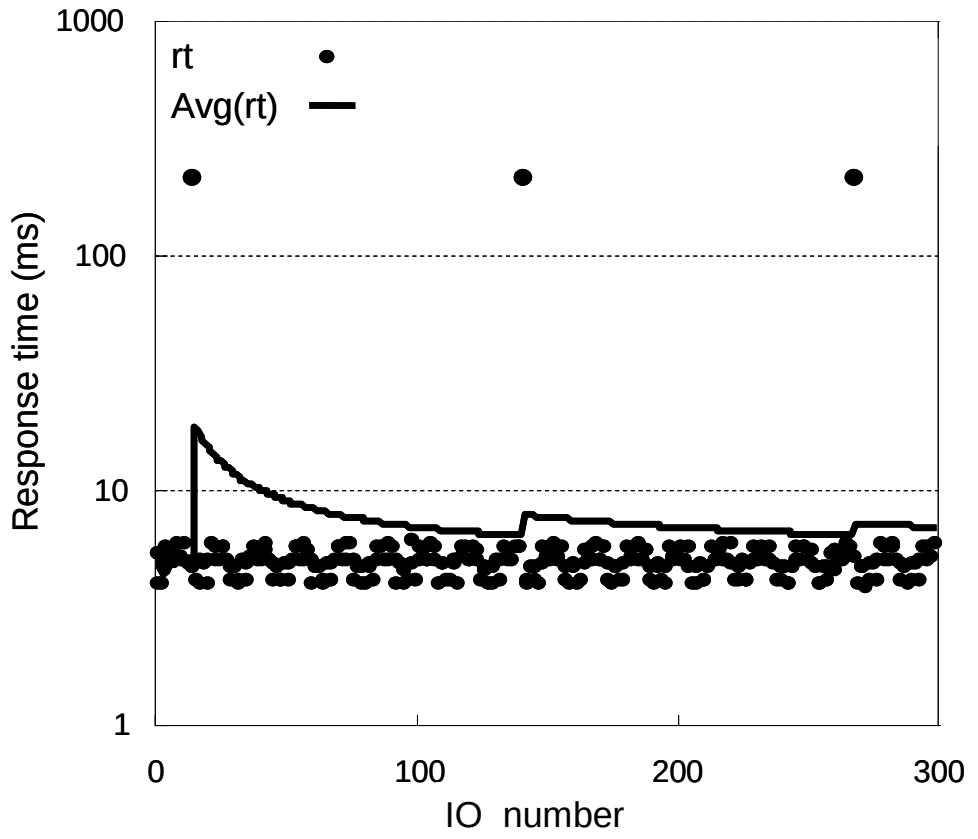

**Figure 4: Running phase for Kingston DTI**

[4] At the time of writing, we were still waiting for the twelfth device, the recently released Flash PCI card from Fusion-IO, advertised as reaching throughput of 600MB/s for random writes. The benchmarking results will be published on uFLIP web-site.

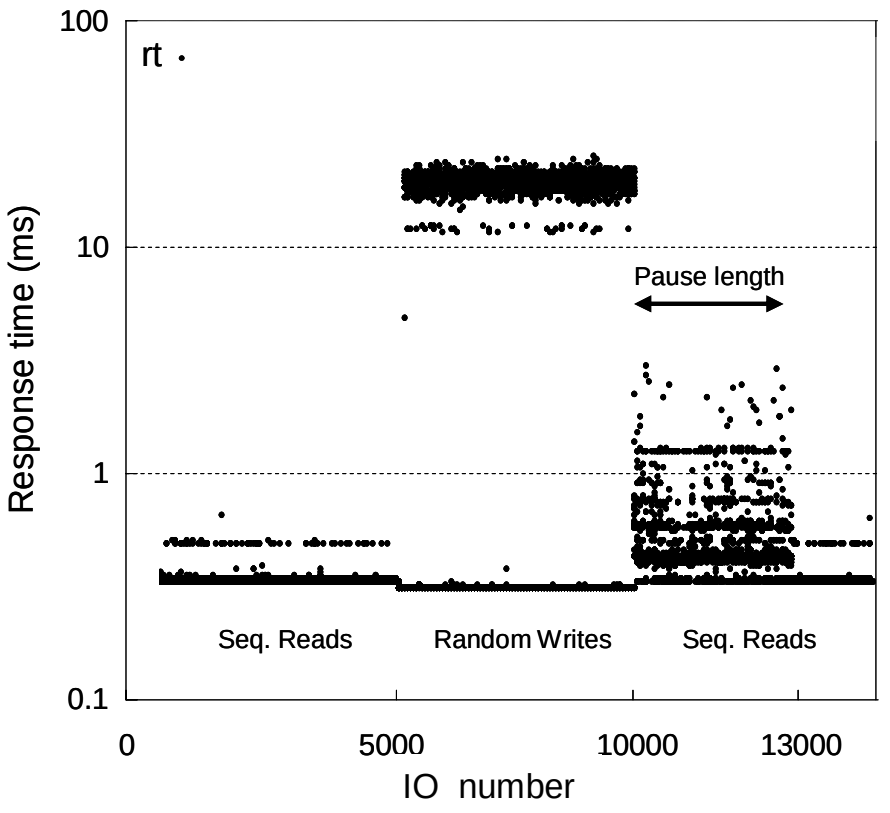

**Figure 5: Pause determination for Mtron**

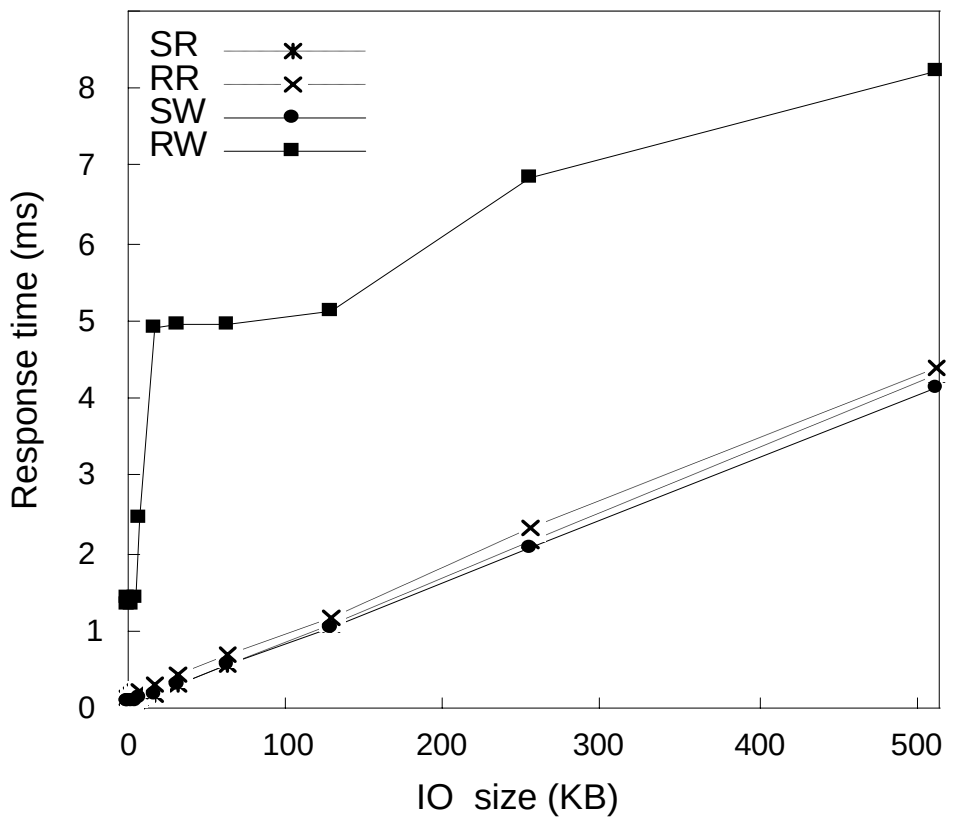

**Figure 6: Granularity for Mtron**

For simplicity, we used the following rules for setting *IOIgnore* and *IOCount*. We set *IOIgnore* = 0 for all devices (no start-up) except the Memoright and Mtron SSDs. For the latter, we used the values 30 and 128, respectively, for experiments involving random writes and 0 for all other experiments. For the SSDs, we set *IOCount* to 1,024 for SR, RR and SW (very small oscillations) and to 5,120 for RW (large oscillations). We set the *IOCount* to 512 in all cases for slow and/or small devices (USB drives, IDE module, and SD card). Note that the values of *IOIgnore* and *IOCount* are automatically scaled by the FlashIO tool when considering mixed workloads.

**Pause between Experiments:** Finally, we must measure the pause required between experiments, by running a pattern of sequential reads, followed by random writes, and sequential reads again. Figure 5 shows the result of this experiment for the Mtron SSD. As before, the *x*-axis shows the time in units of IO operations, while the *y*-axis shows the cost of each operation.

As Figure 5 shows, the lingering effect of the random writes lasts for about 3,000 sequential reads, corresponding to about 2.5 seconds. For this device, we therefore overestimate the pause to 5 seconds.

For all the other devices, including the other SSDs, there was no lingering impact from the random writes. The sequential reads immediately performed as well after the batch of writes as they did before the batch of writes; we therefore set the pause to 1 s (to be conservative).

## 5.2 Benchmark Results

Having set the stage for our benchmarking effort, we now turn to the results of the actual uFLIP micro-benchmarks. As mentioned above, we focus on the results from the seven flash devices indicated in Table 2, as they are very representative for the set. In this section, we cover the most interesting results of our analysis.

**Effect of Granularity:** We first consider the performance on the Granularity micro-benchmark where *IOSize* is varied. We generally expect reads to be cheaper than writes because some writes will generate erase operations, and we also expect random writes to be more expensive than sequential writes as they should generate more erases.

Figure 6 shows the response time (in msec) of each IO operation for the Memoright SSD. Three observations can be made about this figure. First, all reads and sequential writes are very efficient; their response time is linear with a small latency (about 70 µsec for SR/SW and 115 µsec for RR). Second, for rather large random writes, the response time is much higher, at least 5 msec; note that, similar to Figure 3, the cost of random writes alternates between cheap writes (of similar cost to sequential writes) and extremely expensive erase operations (tens of ms). Third, small random writes are serviced much faster; apparently due to caching as four writes of 4KB take about as much time as two writes of 8KB and one write of 16KB.

In comparison, Figure 7 shows the response time for the Kingston DTI USB flash drive. In this figure, the response time of random writes is omitted, as it is a rather constant value around 260 msec. As the figure shows, for this device the cost of sequential writes is affected strongly by the IO granularity, as smaller writes incur a significantly higher cost than writes of 32KB. Comparing the two devices, we observe that while random writes are up to a factor of five times slower than the other operations on the Memoright SSD, they are one or two orders of magnitude slower for the Kingston DTI USB flash drive. This is undoubtedly due to more advanced hardware and FTL on the Memoright SSD (Figure 1 shows that the Memoright SSD includes an FGPA, 16 MB of RAM and a condenser!).

The remaining experiments were run with IO sizes of 32KB. Furthermore, since the performance of reads is excellent, we focus largely on the performance of (random) writes.

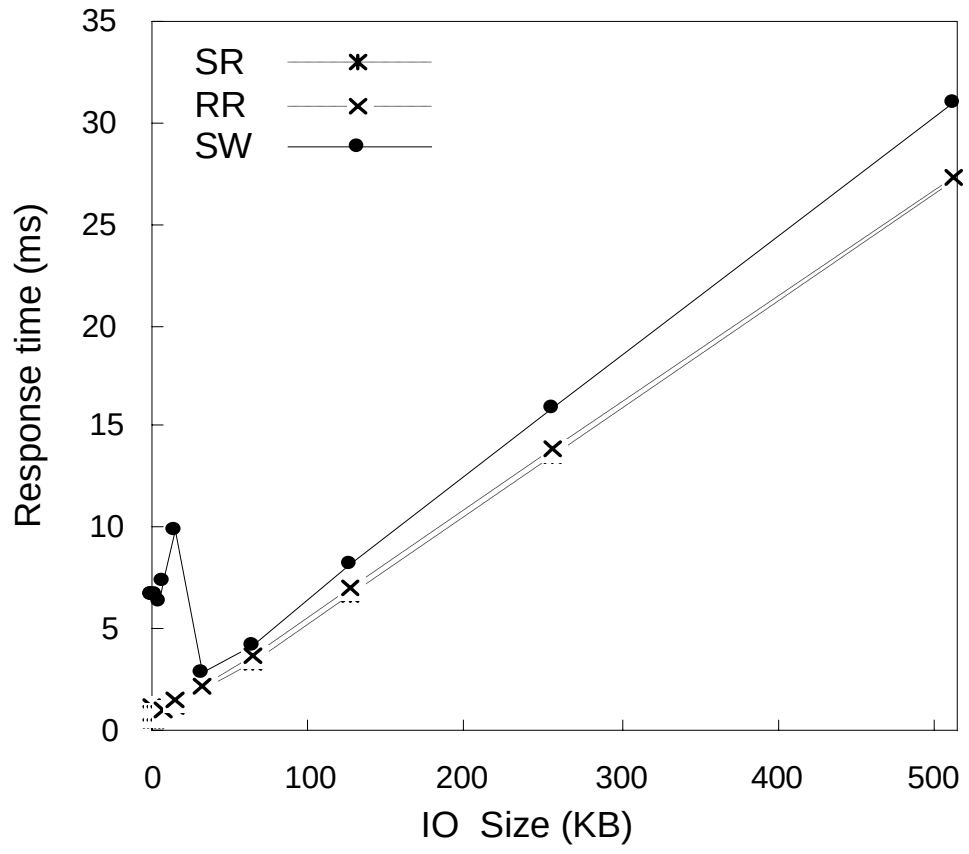

**Figure 7: Granularity for Kingston DTI (SR,RR,SW)**

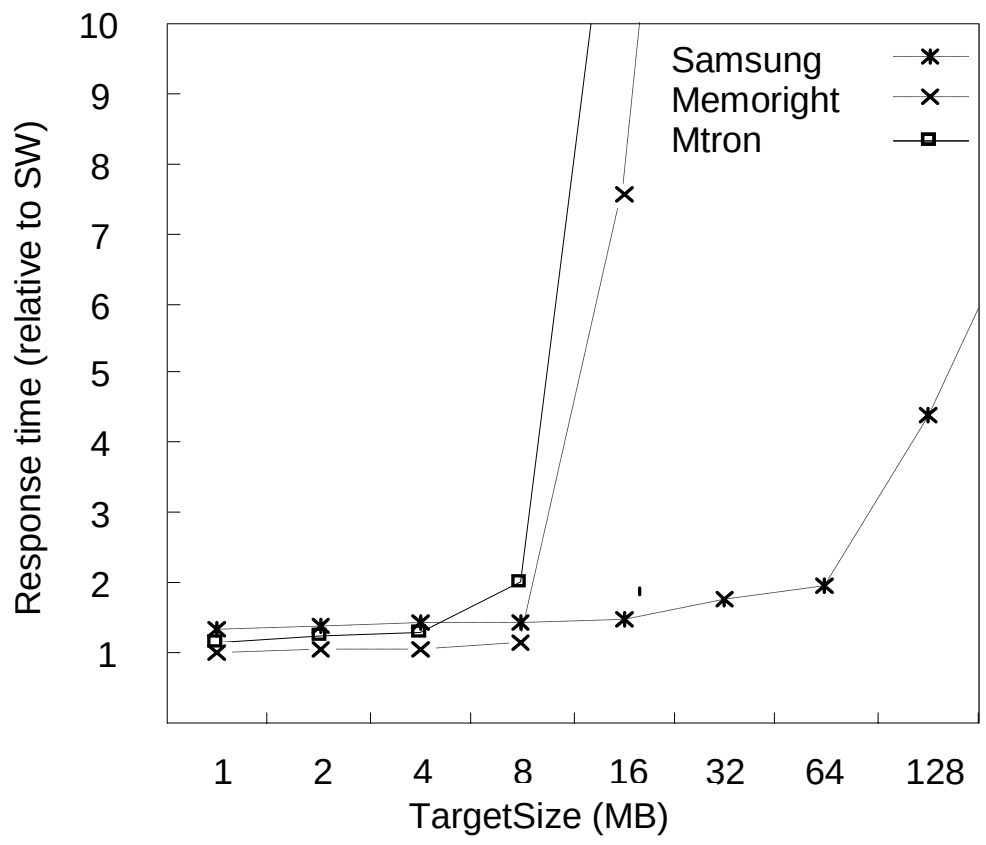

**Figure 8: Locality for Samsung, Memoright and Mtron**

**Effect of Locality:** Figure 8 shows the response time of random writes (relative to sequential writes) as the target size grows from very small (local) to very large (note the logarithmic *x*-axis). Our expectation was that doing random writes within a small area might improve their performance. The figure verifies this intuition, as random writes within a small area have nearly the same response time as sequential writes. The figure shows, however, that the exact effect of locality varies between devices, both in terms of the area that the random writes can cover, and in terms of their relative performance. Note that some low end USB devices (e.g., Kingston DTI) do not show any benefit by focusing random IOs in a reduced area.

Locality does not affect performance of sequential writes, until the area becomes so small that the writes become in-place writes (see below for the effect of in-place writes).

**Key Characteristics:** We now turn our attention to Table 3, which succinctly summarizes some key results from several experiments. In fact, it can be argued that the results in the table describe the key characteristics of the devices, and could be used as the basis for a course classification or categorization. In the following, we will discuss the result columns from left to right.

First, SR, RR, SW, RW indicate the cost of a corresponding IO operation of 32KB. These columns show that there is a large difference in performance between the USB flash drives and the other devices, but also between low-end and high-end SSDs. For the high-end SSDs, even the random write performance is quite good. In fact, as we explore more results, the high-end SSDs distinguish themselves further from the rest.

The fifth column of Table 3 indicates the effect of the Pause micro-benchmark on the random write baseline pattern. No value indicates that this had no effect, which in turn indicates that no asynchronous page reclamation is taking place. For the high-end SSDs, however, inserting a pause improves the performance of the random writes to the point where they behave like sequential writes. Interestingly, however, the length of the pause when that happens is precisely the time required on average for a random write. Thus, no true response time savings are seen by inserting this pause, as the total workload takes the same overall time regardless of the length of the pause. A similar effect is seen with the Burst micro-benchmark.

The sixth column of Table 3 summarizes the effect of Locality on random writes, which we already explored with Figure 8; it shows the size of "locality area" for random writes in MB and, in parentheses, the maximum cost of random writes within that area relative to the average cost for sequential writes.

The seventh column of Table 3 summarizes a similar effect for the Partitioning micro-benchmark. The goal of that experiment was to study whether concurrent sequential write patterns over many partitions degrade the performance of the sequential writes. The column shows the number of concurrent partitions that can be written to without significant degradation of the performance, as well as the cost of the writes relative to sequential writes to a single partition. Note that when writing to more partitions than indicated in this column, the write performance degrades significantly.

The last three columns of Table 3 investigate the Order micro-benchmark. The eighth and ninth columns show the cost of the reverse (*Incr* = –1) and in-place (*Incr* = 0) patterns, respectively, compared to the cost of sequential writes. As the columns show, the effect of the in-place pattern, in particular, varies significantly between devices, ranging from time savings of about 40% for the Samsung SSD, to important performance degradation for the Kingston DTI USB flash drive. The final column shows the impact of large increments (gaps from one 1 MB to 8 MB) compared to the cost of random writes. As the column shows, for high end SSDs and for the Transcend IDE Module, the cost is twice or four times the cost of a random writes.

**Table 3: Result summary**

| | **Basic patterns** | | | | **Pause** | **Locality** | **Partitioning** | **Ordered** | | |
|---|---|---|---|---|---|---|---|---|---|---|
| **Device** | SR (ms) | RR (ms) | SW (ms) | RW (ms) | RW (ms) | RW (MB) | RW (Partitions) | Reverse (*Incr* = -1) | In-Place (*Incr* = 0) | Large *Incr* |
| **Memoright** | 0.3 | 0.4 | 0.3 | 5 | 5 | 8 (=) | 8 (=) | = | = | x4 |
| **Mtron** | 0.4 | 0.5 | 0.4 | 9 | 9 | 8 (x2) | 4 (x1.5) | = | = | x2 |
| **Samsung** | 0.5 | 0,5 | 0.6 | 18 | | 16 (x1.5) | 4 (x2) | x1.5 | x0.6 | x2 |
| **Transcend Module** | 1.2 | 1.3 | 1.7 | 18 | | 4 (x2) | 4 (x2) | x3 | x2 | x2 |
| **Transcend MLC** | 1.4 | 3.0 | 2.6 | 233 | | 4 (=) | 4 (x2) | x2 | x2 | x1 |
| **Kingston DTHX** | 1.3 | 1.5 | 1.8 | 270 | | 16 (x20) | 8 (x20) | x7 | x6 | x1 |
| **Kingston DTI** | 1.9 | 2.2 | 2.9 | 256 | | No | 4 (x5) | x8 | x40 | x1 |

**Other Results:** To give a short outline of the results of the remaining micro-benchmarks, which we have not covered in detail, we observed the following:

- Unaligned IO requests result in significant performance degradation for some devices. For instance, on the Samsung SSD, random IOs should be aligned to 16 KB, as otherwise the response time increases from 18 msec to 32 msec.
- The Mix patterns did not affect significantly the overall cost of the workloads. This behavior is very different from hard disks, where combinations of workloads significantly affect their performance.
- Finally, we did not observe any performance improvements from submitting IOs in parallel. In fact, parallel execution with a high degree can cause multiple sequential write patterns to degenerate to random write patterns (more precisely to partitioned write patterns), with the corresponding increase in cost.

## 5.3 Discussion

The goal of the uFLIP benchmark is to facilitate understanding of the behavior of flash devices, in order to improve algorithm and system design against such devices. In this section we have used the uFLIP benchmark to explore the characteristics of a large set of representative devices. From our results, we draw three major conclusions.

First, we have found that with the current crop of flash devices, their performance characteristics can be captured quite succinctly with a small number of performance indicators shown in Table 3.

Second, we observe that the performance difference between the high-end SSDs and the remainder of the devices, including low-end SSDs, is very significant. Not only is their performance better with the basic IO patterns, but they also cope better with unusual patterns, such as the reverse and in-place patterns. Unfortunately, the price label is not always indicative of relative performance, and therefore designers of high-performance systems should carefully choose their flash devices.

Finally, based on our results, we are able give the following design hints for algorithm and system designers:

*Hint 1: Flash devices do incur latency*. Despite the absence of mechanical parts, the software layers incur some overhead per IO operation. Therefore, larger IOs are generally beneficial, even for read operations.

*Hint 2: Block* size *should (currently) be 32KB*. Based on the first hint, large block sizes are beneficial for writes, while an application of the famed five minute rule [4] says 4KB pages are beneficial for reads, based on prices and capacities of the high-end devices we studied. We therefore believe that 32BK is a good trade-off for those high-end devices.

*Hint 3: Blocks should be aligned to flash pages.* This is not unexpected, based on flash characteristics, but we have observed that the penalty paid for lack of alignment is quite severe.

*Hint 4: Random writes should be limited to a focused area.* Our experiments show that, for most devices, random writes to an area of 4–16MB perform nearly as well as sequential writes. Random writes to larger areas are typically expensive and should be avoided; again, however, the high-end SSDs perform much better in this regard.

*Hint 5: Sequential writes should be limited to a few partitions.* Concurrent sequential writes to 4–8 different partitions are acceptable; beyond that performance degrades to random writes.

*Hint 6: Combining a limited number of patterns is acceptable.* In the same vein, concurrent access from a few patterns does not appear to affect the performance of the individual patterns.

*Hint 7: Neither concurrent nor delayed IOs improve the performance.* Due to the absence of mechanical components, IO scheduling is not improved through abundance of pending asynchronous IOs. Furthermore, introducing pauses does not affect total response time.

# 6. CONCLUSION

The design of algorithms and systems using flash devices as secondary storage should be grounded in a comprehensive understanding of their performance characteristics. We believe that the investigation of flash device behavior deserves strong and continuous effort from the community: uFLIP and its associated benchmarking methodology should help define a stable foundation for measuring flash device performance. By making available online (at www.uflip.org) the benchmark specification, the software we developed to run the benchmark, and the results we obtained on eleven devices, our plan is to gather comments and feedback from researchers and practitioners interested in the potential of flash devices.

There are many avenues for future work. First, we would like to facilitate benchmarking efforts, e.g., through (semi-)automatic tuning of experiment length to ensure that the start-up period is omitted and the running phase captured sufficiently well to guarantee given bounds for the confidence interval, while minimizing the IOs issued. Similarly, (semi-)automatic methods for generating benchmark plans would be useful. Second, we are already working on a visualization interface to facilitate the analysis of benchmark results. Third, we plan to improve the uFLIP web-site, to allow the research community to submit benchmark results. And last, but not least, there are many opportunities for using the knowledge gained from the uFLIP benchmark for algorithm and system design for flash devices.

# 7. ACKNOWLEDGMENTS

This research was done in the context of the Mana project (http://mana.escience.dk) funded by the Danish strategic research council and partially supported by the French National Agency for Research (ANR) under RNTL grant PlugDB.